\documentclass{article}
\usepackage[ansinew]{inputenc}
\usepackage{latexsym}
\newcommand{\be}{\begin{equation}}
\newcommand{\ee}{\end{equation}}
\newcommand{\dt}{\mbox{\boldmath$:$}}
\newcommand{\vdt}{\mbox{\bf$\vdots$}}

\newcommand{\no}{\noindent}

\newcommand{\opsigma}{\mbox{\boldmath$\sigma$}}
\newcommand{\opmu}{\mbox{\boldmath$\mu$}}
\newcommand{\bxi}{\mbox{\boldmath$\xi$}}
\newcommand{\bphi}{\mbox{\boldmath$\phi$}}
\newcommand{\bfeta}{\mbox{\boldmath$\widetilde \eta$}}
\begin{document}

\title{Bosonized Quantum Hamiltonian of the Two-Dimensional Derivative-Coupling Model}

\author{\small{L. V. Belvedere $^1$ and A. F. Rodrigues $^2$\,\footnote{Present address: Centro Brasileiro de Pesquisas F\'{\i}sicas (CBPF), Rua Dr. Xavier Sigaud $150$, CEP $22290-180$, Urca, Rio de Janeiro, Brasil.}}\\
\small{\it{Instituto de F\'{\i}sica - Universidade Federal Fluminense}}\\
\small{\it{Av. Litor\^anea S/N, Boa Viagem, Niter\'oi, CEP 24210-340}}\\
\small{\it{Rio de Janeiro, Brasil}}\\
\small{$1)$\,\it{belve@if.uff.br}}\\
\small{$2)$\,\it{armflavio@if.uff.br}}}
\date{\today}
\maketitle

\begin{abstract}
Using the operator formulation we discuss the bosonization of the two-dimensional derivative-coupling model. The fully bosonized quantum Hamiltonian is obtained by computing the composite operators as the leading terms in the Wilson short distance expansion for the operator products at the same point. In addition, the quantum Hamiltonian contains topological terms which give trivial contributions to the equations of motion. Taking into account the quantum corrections to the bosonic equations of motion and to the scale dimension  of the Fermi field operator, the operator solution is obtained in terms of a generalized Mandelstam soliton operator with continuous Lorentz spin (generalized statistics).  
\end{abstract}


\section{Introduction}

\setcounter{equation}{0}

In the past two-dimensional quantum field models with derivative couplings (DC) have been the subject of various investigations within different approaches \cite{Schr,RS,3,4,5,6,7,8}. In \cite{Schr} the two-dimensional model of a massive Fermi field interacting with a massless scalar field via vector-current-scalar derivative coupling was discussed (Schroer model). In \cite{RS} the model of a massless Fermi field interacting with a massive pseudoscalar field via axial-current-pseudoscalar derivative coupling was analyzed (Rothe-Stamatescu model). The so-called DC model is a generalization of the models introduced in Refs. \cite{Schr,RS} and is the theory of a massive Fermi field interacting with both massless pseudoscalar and scalar fields via derivative couplings. 

The ``bosonization'' of fermions has proven to be a very useful technique for solving two-dimensional quantum field models, as well as to obtain non-perturbative informations of non exactly solvable models \cite{W,Col,M,AAR,Swieca}. Within the bosonization approach, the existence of a hidden Thirring interaction \cite{Th} in the DC model has been discussed in Refs. \cite{3,4,6,7,8,9,10}. In Ref. \cite{10}, the model of a massless pseudoscalar field interacting via axial-current-pseudoscalar derivative coupling with a massive Fermi field (modified Rothe-Stamatescu model (MRS model)) has been analyzed. It was shown that the MRS model is equivalent to the Thirring model with an additional derivative interaction. The fully bosonized version of the model is presented. The bosonized composite operators of the quantum Hamiltonian are computed as the leading operators in the Wilson short-distance expansion \cite{Wilson} for the operator product at the same point.

The main purpose of the present work is to follow the approach given in Ref. \cite{10} to compute the bosonized quantum Hamiltonian of the DC model and the quantum corrections to the operator solution. This streamlines the presentations of Refs. \cite{9,10}. The paper is organized as follows: In order to have a self-contained discussion, in Section $2$ we present a brief review of the operator solution of the DC model as presented in Ref. \cite{9}. In Section $3$, using the Wilson short-distance expansion for the operator product at the same point, the fully bosonized quantum Hamiltonian is obtained and the Thirring interaction is displayed. In Section $4$, using the Hamilton's equations of motion, the classical Lagrangian of the bosonized theory is obtained. In Section $5$, taking into account the quantum corrections to the bosonic equations of motion and to the scale dimension of the Fermi field operator, the operator solution is written in terms of a generalized Mandelstam soliton operator with generalized statistics (continuous Lorentz spin). The operator solution of the Schroer model \cite{Schr} is given in terms of a free massive Fermi field with anomalous scale dimension. The concluding remarks are presented in Section $6$.

\section{Operator solution: a brief review}
\setcounter{equation}{0}

The classical Lagrangian density defining the two-dimensional derivative-coupling model of a massive Fermi field 
interacting with two massless Bose fields is given by \footnote{The conventions used are: 
$$ \gamma^0 = \pmatrix{0 & 1 \cr 1 & 0}\,\,,\,\,\gamma^1 = \pmatrix{0 & 1 \cr -1 & 0}\,\,,\,\,\gamma^5 = \gamma^0 \gamma^1\,,\,\epsilon_{01} = 1\,,\,\gamma^\mu \gamma^5\,=\,\epsilon^{\mu \nu}\,\gamma_\nu\,.
$$
$$ g^{00} = - g^{11} = 1\,,x^\pm = x^0 \pm x^1\,,\partial_\pm = \partial_0 \pm \partial_1\,.$$}, 

$$
{\cal L} (x) = \bar \psi (x)\, i \gamma^\mu\,\partial_\mu\,\psi (x)\,
+\,\frac{1}{2}\,\partial_\mu\,\eta (x)\,\partial^\mu\,\eta (x)\,
+\,\frac{1}{2}\,\partial_\mu\,\widetilde \phi (x)\,\partial^\mu\,\widetilde \phi (x)
$$

\be\label{L1}
+\,g\,\Big (\,\bar \psi (x)\,\gamma^\mu\,\psi (x)\,\Big )\,\partial_\mu\,\eta (x)\
+\,\tilde g\,\Big (\,\bar \psi (x)\,\gamma^\mu\,\gamma^5\,\psi (x)\,\Big )\,\partial_\mu\,\widetilde \phi (x)\,-\,m_o\,\bar\psi (x)\,\psi (x),
\ee

\no where $\eta (x)$ is a {\it scalar} field and $\widetilde \phi (x)$ is a {\it pseudoscalar} field. For $g = 0$, except by the presence of a decoupled massless boson field, the model  corresponds to the Rothe-Stamatescu model \cite{RS} in the zero mass limit of the pseudoscalar field $\widetilde\phi$ and modified to include a mass term for the fermion field (modified RS (mRS) model  \cite{10}), and for $\tilde g = 0$ it corresponds to the Schroer model \cite{Schr}. The quantum equations of motion for the fields $\eta$ and $\widetilde\phi$ are 
   
\be\label{eme}
\mbox{\large$\Box$}  \eta (x)\,=\,-\,g\,\partial_\mu
\vdt\,\Big (\,\bar \psi (x)\,\gamma^\mu\,\psi (x)\,\Big )\,\vdt\,= 0\,,
\ee

\be\label{emp}
\mbox{\large$\Box$} \widetilde \phi (x)\,=\,-\,\tilde g\,\partial_\mu\,
\vdt\,\Big (\,\bar \psi (x)\,\gamma^\mu\,\gamma^5\,\psi (x)\,\Big )\,\vdt\,.
\ee

\no  The  notation $\vdt ( \bullet ) \vdt$ in Eqs.(\ref{eme}) and (\ref{emp}) means that the current is computed as the leading term in the Wilson short distance expansion \cite{10,Wilson}.  Due to the conservation of the vector current in Eq.(\ref{eme}) the scalar field $\eta$ is free and massless. For massive fermions ($m_o \neq 0$) the pseudoscalar field $\widetilde\phi$ does not remain free due to the non-conservation of the axial current in Eq.(\ref{emp}),

\be
\mbox{\large$\Box$} \widetilde\phi (x)\,=\,i\,\tilde g\,m_o\,\vdt\,\Big ( \bar \psi (x)\,\gamma^5\,\psi (x) \Big )\,\vdt\,.
\ee

For massless Fermi fields the model described by the Lagrangian (\ref{L1}) is a
scale invariant theory with anomalous scaling dimension \cite{RS}. As in the standard Thirring model \cite{Swieca}, in order to the theory described by the Lagrangian (\ref{L1}) could have the model with a massless fermion as the short distance fixed point, the scale dimension of the mass operator must be 

\be\label{dim}
D_{_{\bar \psi \psi}} < 2\,.
\ee

\no In what follows the mass term should be understood as a perturbation in the scale invariant model \cite{9,10}. 

The operator solution for the quantum equations of motion is given in terms of Wick-ordered exponentials \cite{RS,3,9,AAR},

\be\label{psi}
\psi (x)\,=\,{\cal Z}_\psi^{ -\,\frac{1}{2}} \,\dt\,e^{\,\textstyle i\,[\,g\,\eta (x)\,+\,\tilde g\,\gamma^5\,\widetilde \phi (x)\,]}\,\dt\,\psi^{(0)} (x)\,,
\ee

\no where ${\cal Z}_\psi$ is a wave function renormalization constant \cite{RS,9,10,AAR} and $\psi^{(0)} (x)$ is the free massive Fermi field. The bosonized expression for the free massive Fermi field is given by the Mandelstam field operator \cite{M},

\be
\psi^{(0)} (x) = \Big (\frac{\mu}{2 \pi} \Big )^{1/2}\,e^{\,\textstyle -\,i\,\frac{\pi}{4}\,\gamma^5}\,
\dt\,e^{\textstyle\,i\,\sqrt \pi\,\{\,\gamma^5\,\widetilde \varphi (x)\,+\,\displaystyle\int_{x^1}^{\infty}\,
\partial_0 {\widetilde\varphi} (x^0,z^1)\,d z^1\,\}}\,\dt\,,
\ee

\no where $\mu$ is an infrared regulator reminiscent of the free massless theory. For $m_o = 0$ use can be made of the fact that 

\be
\epsilon_{\mu \nu} \partial^\nu \widetilde \varphi = \partial_\mu \varphi\,.
\ee

\no The meaning of the notation $\dt ( \bullet ) \dt$ in the field operators is that the Wick ordering is performed by a point-splitting limit in which the singularities subtracted are those of the free theory. In this way, the  Wilson short distance expansions are performed using the two-point function of the free massless scalar field \cite{9,10,AAR}

\be
[ \Phi^{(+)} (x) , \Phi^{(-)} (0)]_{_{x \approx 0}} =\,-\,\frac{1}{4 \pi}\,\ln \{- \mu^2 (x^2 \,+\, i\, \epsilon\, x^0)\}\,.
\ee

\no We shall ignore the infrared problems of the two-dimensional massless free scalar field, since the selection rules carried by the Wick-ordered exponentials ensure the construction of a positive metric Hilbert subspace \cite{Swieca,W} for the fermionic sector of the model.

The vector current is computed with the regularized point-splitting limit procedure

\be\label{cur}
{\cal J}^\mu (x) = \lim_{\varepsilon \rightarrow 0}\,\Big \{\,\bar \psi ( x + \varepsilon ) \gamma^\mu\,
e^{\textstyle\,-\,i\,\displaystyle\int_x^{x + \varepsilon }\,
[ g\,\gamma^5\,\epsilon_{\mu \nu}\,\partial\,^\nu \eta (z)\,+\,\tilde g\,\epsilon_{\mu \nu}\,\partial\,^\nu\,\widetilde \phi (z) ]\,d z^\mu}\,\psi (x)\,
-\,V. E. V.\,\Big \}\,,
\ee

\no with the wave function renormalization constant given by

\be
{\cal Z}_\psi (\epsilon)\,=\,e^{\,\{\, \tilde g\,^2 [ \widetilde\phi^{(+)} (x + \epsilon ) , \widetilde\phi^{(-)} (x) ] + g^2 [ \eta^{(+)} (x + \epsilon ) , \eta^{(-)} (x) ]\,\}}\,=\,\Big (\,-\,\mu^2\,\epsilon ^2 \,\Big )^{-\,\frac{1}{4 \pi}\,(g^2\,+\,\tilde g^2 )}\,.
\ee

\no  The vector current is given by

\be\label{vc}
{\cal J}^\mu (x) = j^\mu_{f} (x)\,-\,\frac{\tilde g}{\pi}\,\epsilon^{\mu\nu}\partial_\nu\,\widetilde \phi (x)\,-\, \frac{g}{\pi}\,\partial^\mu\,\eta (x)\,,
\ee

\no where $j^\mu_{f} (x)$ is the free fermion current,

\be
j^\mu_f (x) = - \frac{1}{\sqrt \pi}\,\epsilon^{\mu \nu}\,\partial_\nu \widetilde \varphi (x)\,.
\ee

\no The axial current is

\be\label{avc}
{\cal J}_\mu^5 (x)\, =\, \epsilon_{\mu\nu} {\cal J}^\nu (x)\,= \,-\,\partial_\mu\,\Big (\,\frac{1}{\sqrt \pi}\,\widetilde \varphi (x)\,+\,\frac{\tilde g}{\pi}\,\widetilde\phi (x)\,\Big )\, -\, \frac{g}{\pi}\,\epsilon_{\mu \nu}\partial\,^\nu \eta (x)\,.
\ee

\no Using (\ref{vc}) and (\ref{avc}), the bosonized form of the quantum equations of motion  (\ref{eme}) and (\ref{emp}) are 

\be
\label{emeq}
\Big ( 1 - \frac{g^2}{\pi} \Big )\,\mbox{\large$\Box$} \eta (x)\,=\,0\,,
\ee

\be\label{bem}
\Big ( 1 - \frac{\tilde{g}\,^2}{\pi} \Big )\,\mbox{\large$\Box$} \widetilde \phi (x)\,=\,\frac{\tilde g}{\sqrt \pi}\,
\mbox{\large$\Box$} \widetilde \varphi (x)\,.
\ee

\no The bosonized mass operator takes the form 

\be\label{bmo}
\vdt \Big ( \bar \psi (x)\,\psi (x) \Big ) \vdt\,=\,-\,\frac{\mu}{\pi}\,\dt \cos \,\Big (\,2 \sqrt \pi\,
\widetilde \varphi (x)\,+\,2\, \tilde g\,\widetilde \phi (x)\,\Big ) \dt\,,
\ee

\no and the $\gamma^5$-invariance breaking term is given by,

\be
\vdt \Big ( \bar \psi (x)\,\gamma^5\,\psi (x) \Big ) \vdt\,=\,i\,\frac{\mu}{\pi}\,\dt \sin \,\Big (\,2 \sqrt \pi\,
\widetilde \varphi (x)\,+\,2\, \tilde g\,\widetilde \phi (x)\,\Big ) \dt\,.
\ee

\no For $m_o = 0$ the axial current (\ref{avc}) is conserved. In this case the 
pseudoscalar fields $\widetilde \varphi$ and $\widetilde \phi$ are both free and massless. Notice that, due to the conservation of the vector current (\ref{vc}), the scalar field $\eta$ is a free  massless field, even for $m_o \neq 0$. From the bosonized mass operator (\ref{bmo}) and from the equation of motion (\ref{bem}) we see that for $m_o \neq 0$ the fields $\widetilde\varphi$ and $\widetilde\phi$ are coupled sine-Gordon-like fields.  The mass operator is independent of the scalar (free) field $\eta$ associated with the vector-current-scalar-derivative interaction (Schroer model) in the Lagrangian (\ref{L1}). 

Let us introduce the canonical fields $\eta\,^\prime$ and $\widetilde\phi ^\prime$ by the fields scaling

\be
\Big ( 1 - \frac{g^2}{\pi} \Big ) \eta \,=\,\eta ^\prime\,,
\ee

\be
\Big ( 1 - \frac{\tilde g^2}{\pi} \Big ) \widetilde\phi \,=\,\widetilde\phi ^\prime\,.
\ee

\no For $\tilde g \neq 0$, performing the canonical field transformation

\be\label{cta}
\beta \widetilde\Phi\,=\,2 \sqrt \pi \widetilde\varphi\,+\,\frac{2 \tilde g}{\sqrt{1\,-\,\frac{\tilde g^2}{\pi}}}\,\widetilde\phi^\prime\,,
\ee

\be\label{ctb}
\beta \widetilde\xi\,=\,\frac{2 \tilde g}{\sqrt{1\,-\,\frac{\tilde g^2}{\pi}}}\,\widetilde\varphi\,-\,\,2 \sqrt \pi \widetilde\phi^\prime\,,
\ee

\no with

\be
\beta^2\,=\,\frac{4 \pi}{1\,-\,\frac{\tilde g^2}{\pi}}\,,
\ee

\no the mass operator (\ref{bmo}) is given by

\be
\vdt \Big ( \bar \psi (x)\,\psi (x) \Big ) \vdt\,=\,-\,\frac{\mu}{\pi}\,\dt \cos \,\big (\,\beta\,\widetilde\Phi\,\big ) \dt\,.
\ee

\no  The equation of motion (\ref{bem}) is then reduced to

\be
\mbox{\large$\Box$}\,\widetilde\xi (x) = 0\,.
\ee

\no  The vector current (\ref{vc}) can be rewritten as

\be
{\cal J}_\mu (x) = \,\eta_\mu (x)\,+\,J_\mu^{Th} (x)\,,
\ee

\no where

\be
\eta_\mu (x)\,=\, -\,\frac{g}{\pi}\,\frac{1}{\sqrt{1 - \frac{g^2}{\pi}}}\,\partial_\mu\,\eta (x)\,,
\ee

\no and the Thirring current is given by

\be
J_\mu^{Th} (x)\,=\,-\, \frac{\beta}{2\pi}\,\epsilon_{\mu \nu}\,\partial^\nu\,\widetilde \Phi (x)\,.
\ee

\no As in the case of the MRS model \cite{10}, the field $\xi$ does not contribute to the fermionic current. Using the fact that the field $\xi$ is free and massless 

\be
\varepsilon_{\mu \nu} \partial^\nu \widetilde \xi (x) = \partial_\mu \xi (x)\,,
\ee

\no and using (\ref{cta})-(\ref{ctb}), the Fermi field operator (\ref{psi}) can be rewritten as

\be
\psi (x)\,=\,{\cal Z}_\psi^{-\,\frac{1}{2}}\,\dt\,e^{\textstyle\,i\,[\,\frac{g}{\sqrt{1 - \frac{g^2}{\pi}}}\, \eta^\prime (x) + \,\tilde g \,\xi (x)\,]}\,\dt\,\Psi (x)\,,
\ee

\no where  $\Psi$ is the Fermi field operator of the massive Thirring model given by the Mandelstam operator \cite{M}

\be
\Psi (x) = \Big (\frac{\mu}{2 \pi} \Big )^{1/2}\,
e^{\textstyle\,-\,i\,\frac{\pi}{4}\,\gamma^5}\,
\dt\,e^{\,\textstyle i\,\{\,\gamma^5\,\frac{\beta}{2}\,\widetilde \Phi (x)\,+\,
\frac{2 \pi}{\beta}\,\displaystyle\int_{x^1}^{\infty}\,\partial_0 {\widetilde\Phi} (x^0,z^1)\,d z^1\,\}}\,\dt\,,
\ee

\no with Lorentz spin $S\,=\,\frac{1}{2}$. The Thirring interaction is not affected by the introduction of the vector-current-scalar-derivative coupling corresponding to the Schroer model ($g$-coupling), which implies that the Thirring interaction is a intrinsic property of the MRS model \cite{9}.

\section{Bosonized quantum Hamiltonian}
\setcounter{equation}{0}

In this section we shall obtain the fully bosonized Hamiltonian of the DC model. Following Ref. \cite{10}, the bosonized composite operators of the quantum Hamiltonian are obtained as the leading operators in the Wilson short distance expansion for the operator products at the same point \cite{Wilson}. From the Lagrangian (\ref{L1}), the classical canonical momenta $\pi_{_{\tilde\phi}}$, $\pi_{_{\eta}}$, conjugate to the fields $\tilde\phi$ and $\eta$, are computed by assuming that both classical vector and axial-vector currents are independent of the fields $\widetilde\phi$ and $\eta$. In this way, the momenta are formally given by the expressions

\be\label{cmphi}
\pi_{\tilde\phi} (x)\,=\,\partial^0 {\tilde\phi} (x) + 
\tilde g\, \big (\,\bar\psi (x)\,\gamma^0\,\gamma^5\,\psi (x)\,\big )\,,
\ee

\be\label{cmeta}
\pi_{{\eta}} (x) = \partial^0 {\eta} (x) + 
g\,\big (\, \bar\psi (x)\,\gamma^0\,\psi (x)\,\big )\,.
\ee

\no For $m_o = 0$, the quantum Hamiltonian density of the scale invariant model is obtained from the classical Hamiltonian with the classical fields replaced by their quantum operator counterparts, and is given in terms of the normal-ordered operator products 

$$
{\cal H} (x) = \frac{1}{2}\,\dt \Big ( \partial_0 \tilde\phi (x) \Big )^2 \dt \,+\,\frac{1}{2}\,\dt \Big ( \partial_1 \tilde\phi (x) \Big )^2 \dt\,-\,
\frac{1}{2}\,\dt \Big ( \partial_0 \eta (x) \Big )^2 \dt \,+\,\frac{1}{2}\,\dt \Big ( \partial_1 \eta (x) \Big )^2 \dt\,
$$

\be\label{ham}
-\,i\,\vdt \Big ( \bar\psi (x)\,\gamma^1\,\partial_1\,\psi (x)\ \Big ) \vdt\,-\,\tilde g\,\dt\,\Big ( \bar\psi (x)\,\gamma^1\,\gamma^5\,\psi (x) \Big )\,\partial_1 \tilde\phi (x)\,\dt\,
-\, g\,\dt\,\Big ( \bar\psi (x)\,\gamma^1\,\psi (x) \Big )\,\partial_1 \eta (x)\,\dt\,,
\ee

\no with the  currents given by Eqs. (\ref{vc})-(\ref{avc}). In terms of the spinor components $\psi_\alpha$ ($\alpha = 1,2.$), the kinetic term of the Fermi field in the Hamiltonian (\ref{ham}) can be written as

\be
h (x) =\, - \,i \,\vdt \Big ( \bar\psi (x)\,\gamma^1\,\partial_1\,\psi (x) \Big ) \vdt\,=\,\sum_{\alpha = 1}^2\,( - 1 )^{\alpha + 1}\,h_\alpha (x)\,,
\ee

\no where 

\be\label{halpha}
h_\alpha (x)\,=\,i\,\vdt\,\psi^\dagger_\alpha (x)\,\partial_1\,\psi_\alpha (x)\,\vdt\,.
\ee

\no  We shall compute the composite operator $h_\alpha (x)$ as the leading term in the Wilson short distance expansion for the operator product at the same point using the same regularization as that employed in the computation of the fermionic current (\ref{vc}). To begin with, let us consider the point-splitting limit

\be\label{ih}
h_\alpha (x)\,=\,\lim_{\varepsilon \rightarrow 0}\,\Big \{\,h_\alpha (x ; \varepsilon )\,+\,h. c.\,-\,V. E. V. \Big \}\,,
\ee

\no where $h_\alpha ( x ; \varepsilon )$ is defined by the splitted operator product

\be\label{ha}
h_\alpha (x ; \varepsilon )\,=\,\frac{i}{2}\,\Bigg (\,\dt \psi^\dagger_\alpha (x + \varepsilon )\,
e^{\textstyle\,- i\,\displaystyle\int_{- \infty}^{x + \varepsilon}{\cal A}_\mu^\alpha (z)\,d z^\mu} \dt \,\Bigg )\,\Bigg (\,\dt \,e^{\textstyle\,i\,\displaystyle\int_{- \infty}^{x}{\cal A}_\mu^\alpha  (z)\,d z^\mu}\,\partial_1\,\psi_\alpha (x) \dt\,\Bigg )\,,
\ee

\no with

\be
{\cal A}_\mu^\alpha (z)\,=\,\tilde g\,\epsilon_{\mu \nu}\,\partial ^\nu \widetilde \phi (z)\,+\,g\,\gamma^5_{\alpha \alpha}\,\epsilon_{\mu \nu}\,\partial ^\nu \eta (z)\,.
\ee

\no Using the operator solution (\ref{psi}), the operator product (\ref{ha}) can be written in terms of the Wick-ordered exponentials of the fields $\tilde\phi$ and $\eta$ as 

$$
h_\alpha (x ; \varepsilon )\,=\,{\cal Z}^{-1}_\psi (\varepsilon )\,\Big \{\,\,\dt e^{\,-\,i\,\Sigma_\alpha (x + \varepsilon )} \dt \,
\dt e^{\,i\,\Sigma_\alpha (x)} \dt \,\,h_\alpha^{(0)} ( x ; \varepsilon )
$$

\be\label{ha1}
-\,\frac{1}{2}\,\dt e^{\,-\, i\,\Sigma_\alpha (x + \varepsilon )} \dt\,\Bigg (\,
\dt e^{\,i \,\Sigma_\alpha (x)}\,\Big (\,g\,\partial_1\,\eta (x)\,+\,\tilde g\,\gamma^5_{\alpha\alpha}\, \partial_1 \widetilde\phi (x)\,\Big )\,\dt\,\Bigg )\,\psi_\alpha^{(0)^\dagger} (x + \varepsilon )\psi_\alpha^{(0)}(x) \Big \}\,,
\ee

\no where we have defined

\be
\Sigma_\alpha (x)\,=\,g\,\eta (x)\,+\,\tilde g\,\gamma^5_{\alpha\alpha}\,\widetilde\phi (x)\,+\,\int_{- \infty}^x{\cal A}_\mu^\alpha (z) d z ^\mu\,,
\ee

\no and $h_\alpha^{(0)} ( x ; \varepsilon )$ is the contribution of the kinetic term of the free Fermi field,

\be\label{h0}
h_\alpha^{(0)} ( x ; \varepsilon ) = \frac{i}{2}\,\psi_\alpha^{(0)^\dagger} (x + \varepsilon )\,\partial_1\,\psi_\alpha^{(0)} (x)\,.
\ee

\no In the computation of (\ref{ha1}) we shall use that 

\be\label{u1}
\Big (\,\gamma^5_{\alpha\alpha}\,\varepsilon^\mu\,\partial_\mu\,+\,\varepsilon^\mu\,\epsilon_{\mu \nu} \partial^\nu \Big )\,\tilde\phi (x) = \mp\,\varepsilon^\pm\,\partial_\pm \tilde\phi (x)\,\,\,,\,\,\alpha = 1 , 2\,,
\ee

\no and that, if $ [ B , A ] = $ c - number,

\be\label{AB}
e^{-\,B}\,A\,=\,A\,e^{-\,B}\,-\,[ B , A ]\,e^{-\,B}\,,
\ee

$$
\Big (\,\dt e^{\,-\,i\,a\,\Phi (x)} \dt \,\Big )\,\Big (\, \dt e^{\,i\,a\,\Phi (y)}\,\partial_1\,\Phi (y) \dt \Big )\,=
$$

\be
e^{\,a^2\,D^{(+)} (x - y)}\,\Bigg \{\,\dt e^{\,-\,i\,a\,[\Phi (x)\,-\,\Phi (y)]}\,\partial_1\,\Phi (y) \dt \,-\,
i\,a\,\Big (\,\partial_{y^1}\,D^{(+)} (x - y)\,\Big )\,
\dt e^{\,-\,i\,a\,[\Phi (x)\,-\,\Phi (y)]} \dt\,\Bigg \}\,,
\ee

\no where

\be
D^{(+)} (x) = [ \Phi^{(+)} (x)\,,\,\Phi^{(-)} (0) ]\,.
\ee

\no Performing the normal ordering of the exponentials of the fields $\eta$ and $\tilde\phi$, we can decompose (\ref{ha1}) as follows 

\be\label{h}
h_\alpha (x ; \varepsilon )\,=\,h^{(I)}_\alpha (x ; \varepsilon )\,+\,h^{(II)}_\alpha (x ; \varepsilon )\,+\,h^{(III)}_\alpha (x ; \varepsilon )\,.
\ee

\no where

\be\label{hI}
h^{(I)}_\alpha (x ; \varepsilon )\,=\,\dt e^{\,\pm\,i\,\varepsilon^\pm\,\partial_\pm\,[\,\tilde g \widetilde \phi (x)\,\mp\,g\,\eta (x)]} \dt \,h_\alpha^{(0)} (x ; \varepsilon )\,,
\ee

\be\label{hII}
h^{(II)}_\alpha (x ; \varepsilon )\,=\,-\,\frac{1}{2}\,\Big ( \psi_\alpha^{(0)^\dagger} (x + \varepsilon )\psi_\alpha^{(0)}(x) \Big )\,\dt e^{\,\pm\,i\,\varepsilon^\pm\,\partial_\pm \,[\,\tilde g \widetilde\phi (x)\,\mp\ g \eta (x) ]}\,\partial_1 \big (\,\gamma^5_{\alpha \alpha}\,\tilde g\,\widetilde\phi (x)\,+\,g\,\eta (x) \big ) \dt\,,
\ee

\be\label{hIII}
h^{(III)}_\alpha (x ; \varepsilon )\,=\,\,\frac{i}{2}\,F_\alpha (\varepsilon )\,\Big (
\psi_\alpha^{(0)^\dagger} (x + \varepsilon )\psi_\alpha^{(0)}(x) \Big )\,\dt e^{\,\pm\,i\,\varepsilon^\pm\,\partial_\pm \,[\,\tilde g \widetilde\phi (x)\,\mp\,g \eta (x) ]} \dt \,,
\ee

\no where the singular function $F_\alpha ( \varepsilon )$ is given by 

\be\label{f}
F_\alpha (\varepsilon )\, =\,\pm\,( g^2 + \tilde g^2 )\,\frac{1}{2 \pi \varepsilon^\pm}\,.
\ee

Let us consider the term $h^{(I)}$. Using that \cite{10,AAR}

\be\label{h00}
h_\alpha^{(0)} ( x ; \varepsilon ) =\,\pm\,\frac{1}{8}\,\dt \Big (\,\partial_\pm\,\tilde\varphi (x) \Big )^2 \dt \,\pm\,\frac{1}{4 \pi\,(\varepsilon^\pm )^2}\,+\,{\cal O} (\varepsilon )\,,
\ee

\no  in order to compute the leading operator in 
the $\varepsilon$-expansion of $h^{(I)} ( x ; \varepsilon )$  we need to retain terms up to 
second order in $\varepsilon$ in the exponentials, 

$$
h_\alpha^{(I)} (x) = h^{(I)}_\alpha (x ; \varepsilon ) + h.c. - V.E.V.\,,
$$

\no and combining the contributions from the two spinor components, the operator $h^{(I)} (x)$ is given by

$$
h^{(I)} (x) = h_1^{(I)}(x) - h_2^{(I)} (x)
$$

\be\label{fhI}
= \,{\cal H}^{^{(0)}}_{\tilde \varphi} (x)\,-\,\frac{g^2}{\pi}\,{\cal H}^{^{(0)}}_{\eta} (x)\,-\,
\frac{\tilde g^2}{\pi}\,{\cal H}^{^{(0)}}_{\tilde \phi} (x)\,
+\,\frac{g\,\tilde g}{\pi}\,\big ( \partial_0\,\widetilde \phi (x)\,\partial_1 \eta (x)\,+\,\partial_1 \widetilde \phi (x)\,\partial_0\,\eta (x) \big )\,,
\ee

\no where ${\cal H}^{^{(0)}}_{\Phi} (x)$ is the canonical quantum Hamiltonian density of the free massless field

\be
{\cal H}^{^{(0)}}_{\Phi} (x) \,=\,
 \frac{1}{2}\,\Big \{\,\dt \Big (\partial_0 \Phi (x) \Big )^2 \dt \, +\, \dt \Big (\partial_1 \Phi (x) \Big )^2 \dt \,\Big \}\,.
\ee

\no The first term in (\ref{fhI}) corresponds to the bosonized Hamiltonian of the free massless Fermi field \cite{10,AAR}. The terms proportional to $g^2$ and $\tilde g^2$  in (\ref{fhI}) are quantum corrections to the free Hamiltonian  of the fields $\eta$ and $\tilde\phi$. Defining the current

\be\label{Kmu}
K_\mu\,=\,-\,\big (\,\tilde g\,\epsilon_{\mu \nu} \partial^\nu \widetilde\phi\,+\,g\,\partial_\mu \eta \big )\,,
\ee

\no the contribution (\ref{fhI}) can be rewritten as 

\be\label{ffhI}
h^{(I)} (x) =  \,{\cal H}^{^{(0)}}_{\tilde \varphi} (x)\,-\,\frac{1}{2\pi}\,\Big \{\,\dt\,\Big ( K_0 (x) \Big )^2\,\dt\,+\,\dt\,\Big ( K_1 (x) \Big )^2\,\dt\,\Big \}\,.
\ee

Let us now consider the second term $h^{(II)}$. To this end, we must compute the operator product of the free fermion field appearing in Eq. (\ref{hII}). Using (\ref{u1}) and normal ordering the exponential, one obtains 

\be\label{u2}
\psi^{(0)^\dagger}_\alpha (x + \varepsilon ) \psi^{(0)}_\alpha (x)\,=\,\frac{1}{2 i \pi\,\varepsilon^\pm}\,\dt e^{\,\pm\,i\,\sqrt\pi\,\varepsilon^\pm\,\partial_\pm\,\tilde\varphi (x)} \dt\,.
\ee

\no Introducing 

\be\label{J}
{\cal J}_\pm (x)  = {\cal J}_0 (x) \pm {\cal J}_1 (x)\,=\,\pm\,\partial_\pm\,\Big \{\,\frac{1}{\sqrt\pi}\,\tilde\varphi (x) + \frac{\tilde g}{\pi}\,\widetilde\phi (x)\,\mp\,\frac{g}{\pi}\,\eta (x) \Big \}\,,
\ee

\no and using (\ref{u2}),  we can write (\ref{hII}) as

\be\label{II}
h_\alpha^{(II)} (x ; \varepsilon ) = \,\frac{i}{4 \pi\,\varepsilon^\pm}\,
\dt e^{\,i\,\pi\,\varepsilon^\pm\,{\cal J}_\pm (x)}\,\partial_1 \Big ( g \,\eta (x)\,+\,\tilde g\,\gamma^5_{\alpha \alpha}\,\widetilde\phi (x) \Big ) \dt\,.
\ee

\no Expanding the exponential in (\ref{II}) in powers of $\varepsilon$ up to first order, one has

\be
h_\alpha^{(II)} (x) = h^{(II)}_\alpha (x ; \varepsilon ) + h.c. - V.E.V. =\,\frac{1}{2}\,\dt {\cal J}_\pm (x)\,
\,\partial_1 \Big ( g \,\eta (x)\, +\,\tilde g\, \gamma^5_{\alpha\alpha}\,\widetilde\phi (x) \Big ) \dt \,+\,{\cal O} (\varepsilon )\,.
\ee

\no The leading operator in the second contribution $h^{(II)} (x) = h^{(II)}_1 (x) - h^{(II)}_2 (x)$ is then given by 

\be\label{fhII}
h^{(II)} (x) =\,g\,\dt \Big (\,\bar \psi (x) \gamma^1 \psi (x)\Big )\,\partial_1 \eta (x) \dt \,+
\,\tilde g\,\dt \Big (\,\bar \psi (x) \gamma^1 \gamma^5 \psi (x)\Big )\,\partial_1 \widetilde\phi (x) \dt \,.
\ee

\no As expected from the operator solution (\ref{psi}), the contribution (\ref{fhII}) cancels the corresponding terms in the Hamiltonian (\ref{ham}).

Finally, let us consider the term $h^{(III)}$. Using (\ref{u2}), (\ref{J}) and (\ref{f}), we can write (\ref{hIII}) as follows 

\be
h_\alpha^{(III)} (x ; \varepsilon )\,=\,\pm\,\frac{(g^2 + \tilde g^2 )}{8\pi^2}\,\frac{1}{(\varepsilon^\pm)^2}\,\dt e^{\,i\,\pi\,\varepsilon^\pm\,{\cal J}_\pm (x)} \dt\,.
\ee

\no Expanding the exponential in powers of $\varepsilon$ up to second order, we get

\be
h^{(III)}_\alpha (x) = h_\alpha^{(III)} (x ; \varepsilon ) + h.c. - V.E.V. = \mp\,\frac{1}{8} (g^2 + \tilde g^2 )\,\dt \Big ( {\cal J}_\pm (x) \Big )^2 \dt\,.
\ee

\no The leading term $h^{(III)} (x)$ is given by

\be\label{fhIII}
h^{(III)} (x) \,=\,-\,\frac{1}{4}\,(g^2 + \tilde g^2 )\,\Big \{\,\dt \Big ({\cal J}_0 (x) \Big )^2 \dt\,+\,
\dt \Big ({\cal J}_1 (x) \Big )^2 \dt\,\Big \}\,,
\ee

\no and corresponds to the contribution of the Thirring interaction to the 
quantum Hamiltonian. Collecting all terms (\ref{ffhI})-(\ref{fhII})-(\ref{fhIII}), the bosonized form of the fermionic kinetic term (\ref{ih}) is given by

$$
h (x)\,=\,i\,\vdt \Big ( \bar\psi (x) \gamma^1 \partial_1 \psi (x) \Big ) \vdt\,=
{\cal H}^{^{(0)}}_{\tilde\varphi} (x)\, +\,g\,\dt \Big (\,\bar \psi (x) \gamma^1 \psi (x)\Big )\,\partial_1 \eta (x) \dt
+\,\tilde g\,\dt \Big (\,\bar \psi (x) \gamma^1 \gamma^5 \psi (x)\Big )\,\partial_1 \tilde\phi (x) \dt
$$

\be
-\,\frac{1}{2\pi}\,\Big \{\,\dt\,\Big ( K_0 (x) \Big )^2\,\dt\,+\,\dt\,\Big ( K_1 (x) \Big )^2\,\dt\,\Big \}\,-\,\frac{1}{4}\,(g^2 + \tilde g^2 )\,\,\Big \{\,\dt \Big ( {\cal J}_0 (x) \Big )^2 \dt\,+\,
\dt \Big ({\cal J}_1 (x) \Big )^2 \dt \,\Big \}\,.
\ee

\no Introducing the mass perturbation, the total bosonized quantum Hamiltonian (\ref{ham}) is given by

$$
{\cal H}_{_{bos}} (x) \,=\,
{\cal H}^{^{(0)}}_{\tilde\varphi} (x)\,+\,{\cal H}^{^{(0)}}_\eta (x)\,+\,
{\cal H}^{^{(0)}}_{\tilde\phi}(x)\,-\,\frac{1}{2\pi}\,\Big \{\,\dt\,\Big ( K_0 (x) \Big )^2\,\dt\,+\,\dt\,\Big ( K_1 (x) \Big )^2\,\dt\,\Big \}
$$

\be\label{bosh}
-\,\frac{1}{4}\,(g^2 + \tilde g^2)\,\Big \{\,\dt \big ({\cal J}_0 (x) \big )^2 \dt\,+\,
\dt \big ({\cal J}_1 (x) \big )^2 \dt\,\Big \}\,+\,m^\prime_o\,\dt \cos \Big ( 2 \sqrt \pi\,\tilde\varphi (x) + 2 \tilde g \widetilde\phi (x) \Big ) \dt\,.
\ee

\no where $m^\prime_o = \mu\,m_o / \pi$. (In what follows we shall supress the subscript ``bos''.)

Before proceeding, we should like to do some formal manipulations which, later, will give us a basis to obtain the bosonized Lagrangian. To begin with, let us write the bosonized Hamiltonian explicitly in terms of the fields $\eta, \widetilde\varphi$ and $\widetilde\phi$.  Decomposing the vector current ${\cal J}_\mu$ as

\be
{\cal J}_\mu\,=\,-\,\epsilon_{\mu \nu} \partial^\nu \,\tilde{{J}}\,-\,\frac{g}{\pi}\,\partial_\mu \eta\,,
\ee

\no where

\be\label{jmu}
\tilde{{J}}\,=\, \frac{1}{\sqrt \pi}\,\widetilde\varphi\,+\,\frac{\tilde g}{\pi}\,\widetilde\phi \,,
\ee

\no and using (\ref{Kmu}), one finds from (\ref{bosh}) ($G^2\,=\,(g^2 + \tilde g^2)/ 2$)

$$
{\cal H}\,=\,\gamma^2\,{\cal H}^{^{(0)}}_{\eta}\,+\,\Big (\,1\,-\,\frac{G\,^2}{\pi}\,\Big )\,{\cal H}^{(0)}_{\tilde\varphi}\,+\,\Big (\,1\,-\,\frac{\tilde g^2}{\pi}\,\Big [\,1\,+\,\frac{G\,^2}{\pi}\,\Big ]\,\Big )\,
{\cal H}^{(0)}_{\tilde\phi}
$$

\be\label{bh1}
\,-\,\frac{\tilde g}{\sqrt \pi}\,\frac{G\,^2}{\pi}\,\Big ( \partial_0 \widetilde\varphi\,\partial_0 \widetilde\phi\,+\,\partial_1 \widetilde\varphi\,\partial_1 \widetilde\phi \Big )\,+\,m_o^\prime\,\cos\,
\Big ( 2 \sqrt \pi \widetilde\varphi\,+\,2\,\tilde g\,\widetilde\phi \Big )\,+\,\delta {\cal H}\,,
\ee

\no where

\be
\gamma^2 (g, \tilde g)\,=\,\Bigg ( 1\,-\,\frac{g^2}{\pi}\,\Big ( 1\,+\,\frac{G^2}{\pi} \Big ) \Bigg )\,,
\ee

\no and $\delta {\cal H}$ is a ``topological'' coupling term given by

\be
\delta {\cal H}\,=\,\frac{g \tilde g}{\pi}\,\big ( \partial_0 \eta\, \partial_1 \widetilde\phi\,+\,\partial_0 \widetilde\phi\, \partial_1 \eta \big )\,+\,g\,\frac{G^2}{\pi}\,\big ( \partial_0 \eta\, \partial_1 \tilde{{J}}\,+\,\partial_0 \tilde{{J}}\, \partial_1 \eta \big )\,.
\ee

\no For $\tilde g^2 \neq 0$, in order to decouple the derivative-coupling term among the sine-Gordon fields $\widetilde\varphi$ and $\widetilde\phi$ in (\ref{bh1}), let us perform the field scaling 

\be\label{varphiprime}
\widetilde\varphi\,^\prime\,=\,\Big (\,1\,-\,\frac{G\,^2}{\pi}\,\Big )^{\frac{1}{2}}\,\widetilde\varphi\,,
\ee

\no and introduce the field combination

\be
\widetilde\Sigma\,=\,\widetilde\varphi\,^\prime\,-\,\frac{\tilde g}{\sqrt \pi}\,\displaystyle\frac{\frac{G\,^2}{\pi}}{\sqrt{\Big ( 1 - \frac{G\,^2}{\pi} \Big )}}\,\widetilde\phi\,.
\ee

\no We obtain,

\be\label{bh2}
{\cal H}\,=\,\gamma^2\,{\cal H}^{^{(0)}}_{\eta}\,+\,{\cal H}^{(0)}_{\tilde\Sigma}\,+\,\alpha ^2\,{\cal H}^{(0)}_{\tilde\phi} \,+\,m_o^\prime\,\cos\,
\Bigg ( 2 \sqrt{\frac{\pi}{1 - \frac{G\,^2}{\pi}}} \widetilde\Sigma\,+\,\frac{2\,\tilde g}{1\,-\,\frac{G\,^2}{\pi}}\,\widetilde\phi \Bigg )\,+\,\delta {\cal H}\,,
\ee

\no where

\be
\alpha ^2\,=\frac{1\,-\,\frac{G^2}{\pi}\,-\,\frac{\tilde g ^2}{\pi}}{1\,-\,\frac{G\,^2}{\pi}}\,.
\ee

\no Defining the canonical field $\widetilde\phi^\prime$, 

\be
\widetilde\phi ^\prime\,=\,\alpha\,\widetilde\phi\,,
\ee

\no and introducing the canonical field transformation,

\be\label{ctt1}
{{\mbox{\ss}}} \,\widetilde\Phi\,=\,2\,\sqrt{\frac{\pi}{1\,-\,\frac{G\,^2}{\pi}}}\,\Bigg (\,\widetilde\Sigma\,+\,
\frac{\frac{\tilde g}{\sqrt \pi}}{\sqrt{1\,-\,\frac{G\,^2}{\pi}\,-\frac{\tilde g^2}{\pi}}}\,\widetilde\phi ^\prime\,\Bigg )\,,
\ee

\be\label{ctt2}
{\mbox{\ss}} \,\widetilde\xi\,=\,2\,\sqrt{\frac{\pi}{1\,-\,\frac{G\,^2}{\pi}}}\,\Bigg (\,
\frac{\frac{\tilde g}{\sqrt \pi}}{\sqrt{1\,-\,\frac{G\,^2}{\pi}\,-\frac{\tilde g^2}{\pi}}}\,\widetilde\Sigma\,-\,\widetilde\phi ^\prime\,\Bigg )\,,
\ee

\no where

\be
{\mbox{\ss}} ^2\,=\,\frac{4 \pi}{\Big (\,1 \,-\,\frac{\tilde g^2}{\pi} \Big )\,-\,\frac{G\,^2}{\pi}}\,,
\ee

\no the bosonized Hamiltonian is given by

\be\label{bh3}
{\cal H}\,=\,{\cal H}^{^{(0)}}_{\eta^\prime}\,+\,{\cal H}^{(0)}_{\tilde\xi}\,+\,
{\cal H}^{(0)}_{\tilde\Phi}\,+\,m_o^\prime\,\cos\,\big ( \,{\mbox{\ss}}\,\widetilde\Phi\,\big )\,+\,\delta {\cal H}\,,
\ee

\no where we have defined the canonical field 

\be
\eta^\prime\,=\,\gamma\,\eta\,,
\ee

\no and the topological term $\delta {\cal H}$ can be written as

\be\label{a}
\delta {\cal H}\,=\,-\,a\,\big ( \epsilon_{10} \partial^0 \widetilde\Phi\,\partial_1 \eta ^\prime\,+\,\epsilon_{10} \partial^0 \eta ^\prime\,\partial_1 \widetilde\Phi \big )\,+\,b\,
\big ( \epsilon_{10} \partial^0 \widetilde\xi\,\partial_1 \eta ^\prime\,+\,\epsilon_{10} \partial^0 \eta ^\prime\,\partial_1 \widetilde\xi \big )\,,
\ee

\no with

\be
a\,=\,g\,\gamma^{- 1}\frac{\mbox{\ss}}{2 \pi}\,\Big ( \frac{\tilde g^2}{\pi}\,+\,\frac{G^2}{\pi} \Big )\,\,\,\,,\,\,\,\,b\,=\,\frac{g \tilde g}{\pi}\,\gamma^{- 1}\,.
\ee

\section{Bosonized Lagrangian}
\setcounter{equation}{0}

As in the Thirring model, in order to perform computations within the functional approach, we need to know the fully bosonized Lagrangian. In this section we shall consider the reconstruction of the classical bosonic Lagrangian corresponding to 
the Hamiltonian (\ref{bosh}). To begin with, let us denote by $\Pi$ the new canonical momenta associated with the bosonic fields of the effective bosonized theory described by (\ref{bh3}). The corresponding Lagrangian is given by

\be\label{fbl}
{\cal L} (x)\,=\,\Pi_{\tilde \xi} (x)\,\partial_0 \widetilde\xi (x)\,+
\Pi_{\eta^\prime} (x)\,\partial_0 \eta^\prime (x)\,+\,\Pi_{\tilde \Phi} (x)\,\partial_0 \widetilde\Phi (x)\,-\,
{\cal H} (x)\,.
\ee

\no In order to obtain the momenta $\Pi_\Phi$, we shall consider the Hamilton's equations of motion

\be
\partial_0\,\Pi_\Phi \,=\,-\,\frac{\partial{\cal H}}{\partial \Phi}\,+\, \partial_1\,\Big (\,\frac{\partial {\cal H}}{\partial (\partial_1 \Phi )}\,\Big )\,
\ee

\no and the Lorentz invariance of the model. To begin with, the canonical equation of motion for the momentum $\Pi_{\eta ^\prime}$ is
given by

\be\label{emphi}
\partial_0 \Pi_{\eta^\prime}\,=\,\partial_1^2 \eta ^\prime\,-\,a\,\partial_1 \epsilon_{1 0} \partial^0\,\widetilde\Phi\,
+\,b\,\partial_1 \epsilon_{1 0} \partial^0\,\widetilde\xi\,.
\ee

\no From (\ref{emphi}), the Lorentz invariance of the theory requires that

\be\label{M1}
\Pi_{\eta ^\prime}\,=\,\partial_0 \eta ^\prime\,-\,a\,\epsilon_{0 1} \partial^1\,\widetilde\Phi\,
+\,b\, \epsilon_{0 1} \partial^1\,\widetilde\xi\,,
\ee

\no implying the following equation of motion for the field $\eta^\prime$

\be
\Box\,\eta ^\prime\,=\,a\,\partial^\mu \epsilon_{\mu \nu} \partial^\nu \widetilde\Phi\,-\,b\,\partial^\mu \epsilon_{\mu \nu} \partial^\nu \widetilde\xi\,\equiv\,0\,.
\ee

\no For the momentum $\Pi_{\tilde\xi}$ one finds

\be
\partial_0 \Pi_{\tilde\xi}\,=\,\partial_1^2 \widetilde\xi\,+\,b\,\partial_1 \epsilon_{1 0} \partial^0\,\eta^\prime\,.
\ee

\no The Lorentz invariance requires

\be\label{M2}
\Pi_{\tilde\xi}\,=\,\partial_0 \widetilde\xi\,+\,b\,\epsilon_{0 1} \partial^1\,\eta ^\prime\,,
\ee

\no which leads to the following equation of motion

\be
\Box\,\widetilde\xi\,=\,-\,b\,\partial^\mu \epsilon_{\mu \nu} \partial^\nu \eta ^\prime\,\equiv\,0\,.
\ee

\no For the momentum $\Pi_{\tilde\Phi}$ we have

\be
\partial_0 \Pi_{\tilde\Phi}\,=\,\partial_1^2 \widetilde\Phi\,-\,a\,\partial_1 \epsilon_{1 0} \partial^0\,\eta^\prime\,+\,{\mbox{\ss}}\,m_o^\prime\,\sin \mbox{\ss} \widetilde\Phi\,,
\ee

\no such that

\be\label{M3}
\Pi_{\tilde\Phi}\,=\,\partial_0 \widetilde\Phi\,-\,a\,\epsilon_{0 1} \partial^1\,\eta ^\prime\,,
\ee

\no and thus

\be
\Box\,\widetilde\Phi\,-\,\mbox{\ss} m_o^\prime \sin \mbox{\ss} \widetilde\Phi\,=\,a\,\partial^\mu \epsilon_{\mu \nu} \partial^\nu \eta ^\prime\,\equiv\,0\,.
\ee

\no Now, let us write the momenta (\ref{M1})-(\ref{M3}) as follows

\be
\Pi_{\eta ^\prime}\,=\,\widehat\Pi_{\eta ^\prime}\,+\,\delta \Pi_{\eta ^\prime}\,,
\ee

\be
\Pi_{\tilde\xi}\,=\,\widehat\Pi_{\tilde\xi}\,+\,\delta \Pi_{\tilde\xi}\,,
\ee

\be
\Pi_{\tilde\Phi}\,=\,\widehat\Pi_{\tilde\Phi}\,+\,\delta \Pi_{\tilde\Phi}\,,
\ee

\no where

\be\label{1}
\delta \Pi_{\eta ^\prime}\,=\,-\,a\,\epsilon_{01} \partial^1 \widetilde\Phi\,+\,b\,\epsilon_{01} \partial^1 \widetilde\xi\,,
\ee

\be\label{2}
\delta \Pi_{\tilde\xi}\,=\,b\,\epsilon_{01} \partial^1 \eta ^\prime\,,
\ee

\be\label{3}
\delta \Pi_{\tilde\Phi}\,=\,-\,a\,\epsilon_{01} \partial^1 \eta ^\prime\,,
\ee

\no and $\widehat \Pi$ are the standard canonical momenta of the free theory.  The Lagrangian of the bosonized theory (\ref{fbl}) which takes into account the quantum corrections to the bosonic equations of motion is given by

\be
{\cal L}\,=\,\frac{1}{2}\,(\partial_\mu \eta ^\prime )^2\,+\,
\frac{1}{2}\,(\partial_\mu \widetilde\xi )^2\,+\,\frac{1}{2}\,(\partial_\mu \widetilde\Phi )^2\,-\,m_o^\prime\,\cos \mbox{\ss} \widetilde\phi\,+\,\delta {\cal L}\,,
\ee

\no where

$$
\delta {\cal L}\,=\,\delta \Pi_{\eta ^\prime}\,\partial_0 \eta ^\prime\,+\,\delta \Pi_{\tilde\xi}\,\partial_0 \widetilde\xi\,+\,
\delta \Pi_{\tilde\Phi}\,\partial_0 \widetilde\Phi\,-\,\delta {\cal H}\,
$$

\be
=\,-\,a\,(\epsilon_{\mu \nu} \partial_\nu \widetilde\Phi\,\partial^\mu \eta ^\prime
\,+\,\epsilon_{\mu \nu} \partial_\nu \eta ^\prime\,\partial^\mu \widetilde\Phi )\,+\,b\,
(\epsilon_{\mu \nu} \partial_\nu \widetilde\xi\,\partial^\mu \eta ^\prime
\,+\,\epsilon_{\mu \nu} \partial_\nu \eta ^\prime\,\partial^\mu \widetilde\xi )\,\equiv\,0\,.
\ee

\no This is expected, since from (\ref{a}) and (\ref{2})-(\ref{3}) we have

\be
\delta {\cal H}\,=\,\delta \Pi_{_{\eta^\prime}}\,\partial_0 \eta^\prime\,+\,\delta \Pi_{_{\tilde\xi}}\,\partial_0 \xi\,+\,\delta \Pi_{_{\tilde\Phi}}\,\partial_0 \widetilde\Phi\,.
\ee

\no Defining the trivially conserved topological vector- and pseudo-vector currents

\be
{\bphi}_\mu\,=\,\epsilon_{\mu \nu} \partial^\nu \widetilde \Phi\,,
\ee

\be
{\bxi}_\mu\,=\,\epsilon_{\mu \nu} \partial^\nu \widetilde \xi\,,
\ee

\be
{\bfeta}\,=\,\epsilon_{\mu \nu} \partial^\nu \eta ^\prime\,,
\ee

\no the trivial topological term of the Lagrangian can be written as

\be\label{TL}
\delta {\cal L}\,=\,-\,a\,({\bphi}_\mu\,\partial^\mu \eta ^\prime\,+\,{\bfeta}_\mu\,\partial^\mu \widetilde\Phi )\,+\,b\,
({\bxi}_\mu\,\partial^\mu \eta ^\prime\,+\,{\bfeta}_\mu\,\partial^\mu \widetilde\xi )\,.
\ee

\no The contribution from (\ref{TL}) to the momenta (\ref{M1}), (\ref{M2}) and (\ref{M3}) are obtained by assuming that the topological currents are independent of the Bose fields. Otherwise one gets \footnote{Since the topological contributions to the momenta are given in terms of components of conserved currents, at the quantum level they generate Schwinger terms in the equal-time commutation relations,

$$
\big [ \Pi_{\tilde\xi} (x)\,,\,\Pi_{\tilde\Phi} (y) \big ]\,=\,0\,,
$$

$$
\big [ \Pi_{\eta ^\prime} (x)\,,\,\Pi_{\tilde\xi} (y) \big ]\,=\,2\,i\,b\,\partial_{x^1}\,\delta (x^1\,-\,y^1)\,,
$$

$$
\big [ \Pi_{\eta ^\prime} (x)\,,\,\Pi_{\tilde\Phi} (y) \big ]\,=\,-\,2\,i\,a\,\partial_{x^1}\,\delta (x^1\,-\,y^1)\,.
$$
}

\be
\delta \Pi\,=\,0\,,
\ee

\no in agreement with $\delta {\cal L}\,=\,0$. Notice that for $g = 0$, $\delta {\cal H}\,=\,0$. This is nothing more than a reminiscence of the fact that the momenta (\ref{cmphi}) and (\ref{cmeta}) are formally computed by considering the vector and axial currents as being independent of the Bose fields. For $g \neq 0$, the bosonized expression of the derivative coupling piece of the Lagrangian (\ref{L1}) contains terms of the form $(g \tilde g \epsilon_{\mu \nu} \partial^\nu \eta\,\partial^\mu \widetilde\phi)$ and $(g\,\epsilon_{\mu \nu} \partial^\nu \tilde{{J}}\,\partial_\mu \eta)$,  which gives a identically zero contribution to the equations of motion.

In going over to the bosonized quantum theory, the equations of motion for the Bose fields incorporate the corrections that arise  from the Wilson short-distance expansions, which leads to a new sine-Gordon parameter ${\mbox{\ss}}$. The bosonized theory is described by a free massless scalar field $\eta^\prime$, a free massless pseudo-scalar field $\widetilde\xi$, and the sine-Gordon field $\widetilde\Phi$.

Now, let us specialize to the Schroer model ($\tilde g = 0$). From (\ref{bh1}) the bosonized Hamiltonian is given by

\be
{\cal H}_S\,=\,\gamma^2\,{\cal H}^{^{(0)}}_{\eta}\,+\,\Big (\,1\,-\,\frac{g^2}{ 2 \pi}\,\Big )\,{\cal H}^{(0)}_{\tilde\varphi}\,+\,
m_o^\prime\,\cos\,\Big ( 2 \sqrt \pi \widetilde\varphi\,\Big )\,+\,\delta {\cal H}_S\,,
\ee

\no where

\be
\gamma^2 (g)\,=\,\Bigg ( 1\,-\,\frac{g^2}{\pi}\,\Big ( 1\,+\,\frac{g^2}{2 \pi} \Big ) \Bigg )\,,
\ee

\no and

\be
\delta {\cal H}_S\,=\,\frac{g}{\sqrt \pi}\,\frac{g^2}{ 2\pi}\,\big ( \partial_0 \eta\, \partial_1 \widetilde\varphi\,+\,\partial_0 \widetilde\varphi\, \partial_1 \eta \big )\,.
\ee

\no The bosonized Lagrangian is given by

\be
{\cal L}_{_{S}}\,=\,\gamma^2 (g)\,{\cal L}^{^{(0)}}_{\eta}\,+\,\Big ( 1\,-\,\frac{g^2}{2 \pi}\,\Big ) {\cal L}^{^{(0)}}_{\tilde\varphi}\,-\,m_o^\prime\,\cos\,\big ( \,2\,\sqrt \pi\,\widetilde\varphi\,\big )\,.
\ee

\no Performing the field scaling

\be
\widetilde\varphi\,^\prime\,=\, \Big ( 1\,-\,\frac{g^2}{2 \pi}\,\Big )\,\widetilde\varphi\,,
\ee

\no the bosonized Lagrangian of the Schroer model can be written as

\be\label{lagSchr}
{\cal L}_{_{S}}\,=\,\gamma^2 (g)\,{\cal L}^{^{(0)}}_{\eta}\,+\,{\cal L}^{^{(0)}}_{\tilde\varphi^\prime}\,
-\,m_o^\prime\,\cos\,\big ( \,\beta_S\,\widetilde\varphi\,^\prime\,\big )\,,
\ee

\no where

\be
\beta^2_S\,=\,\frac{4 \pi}{1\,-\,\frac{g^2}{2 \pi}}\,.
\ee

\no As we shall see, and contrary to superficial appearence of the Lagrangian (\ref{lagSchr}), the Schroer model corresponds to a derivative vector-coupling of a free massless scalar field with a free massive Fermi field with anomalous scale dimension.

\section{Field Operators}
\setcounter{equation}{0}

In this section we shall consider the operator solution (\ref{psi})-(\ref{vc}) taking into account the quantum corrections to the bosonic equations of motion. To this end, let us express the operator solution in terms of the Bose fields $\eta\,^\prime$, $\widetilde\xi$ and $\widetilde\Phi$. The vector current (\ref{cur}) can be rewritten as

\be
{\cal J}^\mu\,=\,-\,\frac{g}{\pi}\,\gamma^{- 1}\,\partial^\mu \eta\,^\prime\,+\,{\mathrm{J}}\,^\mu\,,
\ee

\no where 

\be
\mathrm{J}\,^\mu\,=\,-\,\frac{{\mbox{\ss}}}{2 \pi}\,\epsilon^{\mu \nu} \partial_\nu \widetilde\Phi\,.
\ee

\no In the quantum theory, taking into account the quantum corrections to the kinetic term of the field $\eta$ carried by the factor $\gamma ( g , \tilde g )$, after the field scaling $\gamma \,\eta\,=\,\eta ^\prime $ a new wave function renormalization is needed,

\be
{\cal Z}^\prime_\psi (\epsilon )\,\rightarrow\,\big (\,-\,\mu^2\,\epsilon^2\,\big )^{\,-\,\frac{1}{4 \pi}\,\big ( \tilde g^2\,+\,g^2\,\gamma^{\,-\,2}\, \big )}\,=\,{\cal Z}_\psi (\epsilon )\,\big (\,-\,\mu^2\,\epsilon^2\,\big )^{\,-\,f (g , \tilde g)}\,,
\ee

\no with

\be
f ( g , \tilde g )\,=\,\frac{g^2}{4 \pi}\,\Bigg [\,\frac{\frac{g^2}{\pi}\,\Big (\,1\,+\,\frac{1}{2 \pi}\,(g^2 + \tilde g^2)\, \Big )}
{1\,-\,\frac{g^2}{\pi}\,\Big (\,1\,+\,\frac{1}{2 \pi}\,(g^2 + \tilde g^2)\, \Big )}\,\Bigg ]\,.
\ee

\no In this way, the Fermi field operator (\ref{psi}) is given by

\be\label{psibos}
\psi (x)\,=\,\Big ({\cal Z}^\prime_\psi \Big )^{\,-\,\frac{1}{2}} \,\dt\,e^{\,\textstyle i\,[\,g\,\gamma^{ - 1}\,\eta\,^\prime (x)\,+\,\tilde g\,\xi (x)\,]}\,\dt\,\Psi (x)\,,
\ee

\no where $\Psi$ is the generalized Mandelstam field operator,

\be\label{gmo}
\Psi (x)\,=\,\Big (\frac{\mu}{2 \pi} \Big )^{1/2}\,e^{\,\textstyle -\,i\,\frac{\pi}{4}\,\gamma^5} \,\dt\,e^{\textstyle\,i\,\frac{{\mbox{\ss}}}{2}\,\Big (\,\gamma^5 \widetilde\Phi (x)\,+\,\big (\,1\,-\,\frac{\tilde g^2}{\pi} \big )\,\displaystyle\int_{x^1}^\infty \partial_0 \widetilde\Phi (x^0 , z^1 )\,d z ^1\,\Big )}\,\dt\,.
\ee

\no The generalized Mandelstam soliton operator (\ref{gmo}) has continuous Lorentz spin \footnote{As in the standard Thirring model \cite{AAR,MS,TFODO}, the operator (\ref{gmo}) with generalized statistics (``spin''), can be formally written in terms of a product of  order ({$\opsigma$}) and  disorder ({$\opmu$}) operators

$$
\Psi (x)\,=\,{\opsigma} (x)\,{\opmu} (x)\,,
$$

\no where

$$
{\opsigma} (x)\,=\,e^{\,\textstyle i\,\frac{\mbox{\ss}}{2}\,\gamma^5\,\widetilde\Phi (x)}\,,
$$

$$
{\opmu} (x)\,=\,e^{\textstyle\,i\,\frac{{\mbox{\ss}}}{2}\,\big (\,1\,-\,\frac{\tilde g^2}{\pi} \big )\,\displaystyle\int_{x^1}^\infty \partial_0 \widetilde\Phi (x^0 , z^1 )\,d z ^1}\,,
$$

\no which satisfy the equal-time dual algebra

$$
{\opmu} (x)\,{\opsigma} (y)\,=\,{\opsigma} (y)\,{\opmu} (x) \,e^{\,2 \pi\,i\,\gamma^5\,S\,\theta (y^1\,-\,x^1)}\,.
$$

\no The commutation between ${\opsigma}$ and ${\opmu}$ produces a dislocation in the field $\widetilde\Phi$ if ${\opsigma}$ is to the right of ${\opmu}$ and leaves it unchanged otherwise.}

\be
S\,=\,\frac{{\mbox{\ss}} ^2}{8 \pi}\,( 1 - \frac{\tilde g^2}{\pi} )\,,
\ee

\no in the range $\tilde g^2 < \pi$, $\tilde g^2 + G\,^2 < \pi$. The standard Mandelstam field operator with Lorentz spin $1 / 2$ \cite{9,M} can be obtained from (\ref{gmo}) by neglecting the quantum corrections coming from the bosonized Hamiltonian, which formally corresponds to consider $G^2 = 0$ in the sine-Gordon parameter $\mbox{\ss}$, and we obtain \cite{9}

\be\label{mop}
\Psi (x)\,=\,\Big (\frac{\mu}{2 \pi} \Big )^{1/2}\,e^{\,\textstyle -\,i\,\frac{\pi}{4}\,\gamma^5} \,\dt\,e^{\textstyle\,i\,\Big (\,\gamma^5\,\frac{\beta}{2}\, \widetilde\Phi (x)\,+\,\frac{2 \pi}{\beta}\,\displaystyle\int_{x^1}^\infty \partial_0 \widetilde\Phi (x^0 , z^1 )\,d z ^1\,\Big )}\,\dt\,,
\ee

\no with

\be
\beta^2\,=\,\frac{4 \pi}{1\,-\,\frac{\tilde g^2}{\pi}}\,.
\ee

\no In this case, and for $g = 0$, we recover the results of the modified RS model presented in Ref. \cite{9}.

It should be stressed that the Fermi field operator (\ref{psibos}) formally corresponds to the operator solution of the Schroer-Thirring model \cite{9} with two vector-current-scalar derivative couplings. Nevertheless, in the present model, the physical field is the pseudo-scalar field $\widetilde\xi$ that appears in the bosonized Lagrangian, instead of its partner $\xi (x)\,=\,-\,\int_x \epsilon^{\mu \nu} \partial_\nu \widetilde\xi (z) d z ^\mu$. In this way, the Fermi field $\psi$ obeys the quantum equation of motion,

\be\label{qem}
i\,\gamma^\mu \partial_\mu \psi (x)\,=\,-\,g\,\gamma^{ - 1}\,N \big [ \gamma^\mu \psi (x)\,\partial_\mu \eta^\prime (x) \big ]\,-\,\tilde g\,N \big [ \gamma^\mu \gamma^5 \psi (x)\,\partial_\mu \widetilde\xi (x) \big ]\,-\,\tilde g^2\,N \big [ \mathrm{J} _\mu (x) \gamma^\mu \psi (x) \big ]\,+\,{\cal M}\,\psi (x)\,,
\ee

\no where the normal product in (\ref{qem}) is defined by the symmetric limit \cite{RS,9,AAR}

\be
N [ \psi (x) \partial_\mu \Phi (x)]  \doteq \lim_{\varepsilon \rightarrow 0}\,\frac{1}{2}\,\Big \{
\partial_\mu  \Phi (x + \varepsilon ) \psi (x) + \partial_\mu \Phi (x - \varepsilon ) \psi (x) \Big \}\,,
\ee

\no the current $\mathrm{J}_\mu$ is given by (\ref{J}), and ${\cal M}$ is a constant which is infinite, finite, or zero, depending on where the scale dimension of the mass operator 

\be
D\,=\,\frac{{\mbox{\ss}} ^2}{4 \pi}\,,
\ee

\no is greater than, equal to, or less than one \cite{Rothe-Swieca,AAR}. 

From (\ref{gmo}) we read off the Thirring interaction hidden in the derivative-coupling model as an intrinsic property of the axial-current-pseudoscalar derivative interaction (modified Rothe-Stamatescu model) \cite{10}.

Now, let us specialize once more to the case of the Schroer model $\tilde g = 0$. The operator solution (\ref{psi}) can be written as

\be\label{psiSchr}
\psi_{_{S}} (x)\,=\,\Big ( {\cal Z}^\prime_\psi \Big )^{ -\,\frac{1}{2}} \,\dt\,e^{\,\textstyle i\,g\,\gamma^{ - 1}\,\eta\,^\prime (x)}\,\dt\,\Psi^{^{(0)}} (x)\,,
\ee

\no where  $\Psi^{^{(0)}}$ corresponds to a {\it non-canonical free massive Fermi field with continuous Lorentz spin} $S$,

\be
\Psi^{(0)} (x)\,=\,
\Big (\frac{\mu}{2 \pi} \Big )^{1/2}\,
e^{\,\textstyle -\,i\,\frac{\pi}{4}\,\gamma^5}\,
\dt\,e^{\textstyle\,i\,\frac{\beta_{_{S}}}{2}\,\{\,\gamma^5\,\widetilde \varphi ^\prime (x)\,+\,\displaystyle\int_{x^1}^{\infty}\,
\partial_0 {\widetilde\varphi ^\prime\,} (x^0,z^1)\,d z^1\,\}}\,\dt\,,
\ee

\no with the Lorentz spin given by

\be
S\,=\,\frac{\beta_{_{S}}^2}{8 \pi}\,.
\ee

\no The field operator obeys the quantum equation of motion

\be
i\,\gamma^\mu \partial_\mu \psi_{_{S}} (x)\,=\,-\,g\,\gamma^{- 1}\,N \big [ \gamma^\mu \psi_{_{S}} (x)\,\partial_\mu \eta^\prime (x) \big ]\,+\,
{\cal M}\,\psi_{_{S}} (x)\,.
\ee

\no One concludes that in the Schroer model, the scalar field $\eta\,^\prime$, as well as the underlying Fermi field $\Psi^{(0)}$, remain free fields. The quantum effect of the derivative interaction is to give a non-canonical scale dimension for the Fermi field $\Psi^{(0)}$, such that the scale dimension of the mass operator is given by \footnote{The Lehmann spectral function for massive free fermions in the presence of a Wick-ordered exponential of a massless free scalar field has been computed in Ref. \cite{Schr}. The fermion two-point function exhibits the infrared behavior

$$
S_{_{F}} (p)\,\sim\,(p^2\,-\,m^2)^{\,-\,\displaystyle\frac{1}{D}}
$$ 

\no with $D$ given by (\ref{Dmass}).}

\be\label{Dmass}
D\,=\,\frac{1}{1\,-\,\frac{g^2}{2 \pi}}\,.
\ee

\section{Concluding Remarks}
\setcounter{equation}{0}

We have re-examined the bosonization of the two-dimensional derivative coupling model within the operator formulation. By considering the quantum corrections for the bosonic equations of motion, which imply a correction to the scale dimension of the Fermi field 
operator, the operator solution is given in terms of a Mandelstam soliton operator with generalized statistics (``spin''). In order to have a clear interpretation of the generalized Mandelstam operator (\ref{gmo}), let us consider the massless Thirring model, which at the classical level is defined by the equation of motion ($J^\mu = \bar \psi \gamma^\mu \psi$)

\be
i\,\gamma^\mu\,\partial_\mu\,\psi \,=\,-\,\tilde g^2\,J^\mu\,\gamma_\mu \psi \,,
\ee

\no which implies the conservation laws 

\be
\partial_\mu J^\mu\,=\,\epsilon^{\mu \nu} \partial_\mu J_\nu\,=\,0\,.
\ee

\no In this way, the conserved vector current can be written in terms of a scalar potencial $\phi$ as

\be
J^\mu \,=\,-\,\frac{\beta}{2 \pi}\,\epsilon_{\mu \nu} \partial^\nu \widetilde \phi\,=\,
-\,\frac{\beta}{2 \pi}\, \partial_\mu \phi \,.
\ee

\no  A particular class of the one-parameter family of classical solutions for the equation of motion is given by \cite{AAR}

\be\label{cs}
\psi \,=\,e^{\,-\,i\,\tilde g^2\,\frac{\beta}{2\pi}\,\phi }\,\psi^{(0)} \,,
\ee

\no where $\psi^{(0)}$ is the general solution of the free Dirac equation

\be\label{fcs}
\psi^{(0)} \,=\,e^{\,i\,\frac{\beta}{2}\,\big ( \gamma^5 \widetilde\phi \,+\,\phi  \big )}\,.
\ee

\no The value $\beta = 2 \sqrt \pi$ corresponds to the canonical free Dirac field. For

\be\label{beta}
\beta^2\,=\,\frac{4 \pi}{1 - \frac{\tilde g^2}{\pi}}\,,
\ee

\no one can write (\ref{cs}) as

\be
\psi \,=\,e^{\,i\,\big ( \gamma^5 \frac{\beta}{2}\, \widetilde\phi \,+\,\frac{2 \pi}{\beta}\,\phi  \big )}\,.
\ee

\no For massless Fermi fields, the operator (\ref{gmo}) can be written as (unless constant multiplicative factors)

\be\label{qs}  
\Psi\,=\,f (\varepsilon)\,\dt e^{\,-\,i\,\tilde g^2\,\frac{\mbox{\ss}}{2\pi}\,\phi }\dt\,\Psi^{(0)}\,
\ee

\no where $f$ is a renormalization constant and $\Psi^{(0)}$ is the non-canonical free massless Fermi field operator

\be
\Psi^{(0)}\,=\,\dt e^{\,i\,\frac{\mbox{\ss}}{2}\,\big ( \gamma^5 \widetilde\phi \,+\,\phi  \big )}\dt\,.
\ee

\no The operator (\ref{qs}) is nothing more than the quantum version of the classical solution (\ref{cs}). For massive fermions, the scalar field $\phi$ is no longer a free field and, at the quantum level, the operator solution is obtained from (\ref{qs}) by defining  

\be
\phi (x)\,\widehat =\,\int_{x^1}^\infty \partial_0 \widetilde\phi (x^0 , z^1) dz^1\,,
\ee

\no and we get the generalized Mandelstam soliton operator (\ref{gmo}),

\be\label{qsm}
\Psi (x)\,=\,f (\varepsilon)\,\dt e^{\,-\,i\,\tilde g^2\,\frac{\mbox{\ss}}{2\pi}\,{\displaystyle\int_{x^1}^\infty} \partial_0 \widetilde\phi (x^0 , z^1) dz^1
}\dt\,\Psi^{(0)} (x)\,,
\ee

\no where $\Psi^{(0)}$ is the non-canonical free massive Fermi field operator

\be
\Psi^{(0)} (x)\,=\,\dt e^{\,i\,\frac{\mbox{\ss}}{2}\,\big ( \gamma^5 \widetilde\phi \,+\,{\displaystyle\int_{x^1}^\infty }\partial_0 \widetilde\phi (x^0 , z^1) dz^1\,\big )}\dt\,.
\ee

\no The standard Mandelstam operator with Lorentz spin $\frac{1}{2}$ 

\be
\Psi (x)\,=\,
\dt e^{\,i\,\big ( \frac{\beta}{2}\,\gamma^5 \widetilde\phi \,+\,\frac{2 \pi}{\beta}\,{\displaystyle\int_{x^1}^\infty }\partial_0 \widetilde\phi (x^0 , z^1) dz^1\,\big )}\dt\,,
\ee

\no is obtained from (\ref{qsm}) by setting $\mbox{\ss} = \beta$ with $\beta$ given by (\ref{beta}). From (\ref{qsm}) and (\ref{mop}) we read off the presence of the Thirring interaction in the DC model only for $\tilde g^2 \neq 0$.

Let us make a final comment. It is a usual pratice to perform the functional integral bosonization on the partition function level \cite{naive}. It should be stressed that at the partition function level, the free massless Bose fields $\eta ^\prime$ and $\xi$ decouple, which should lead to a misleading conclusion concerning with the equivalence of the DC model with the Thirring model. As rigorously showed in Ref. \cite{fib}, in order to obtain the fermion-boson mapping in the Hilbert space of states, the functional integral bosonization must be performed on the generating functional.

{\bf Acknowledgments}: The authors are greateful to Brazilian Research Council (CNPq) for partial financial support.

\newpage




\end{document}